\documentclass[superscriptaddress,aps,prl,reprint,showpacs]{revtex4-1}

\usepackage[english]{babel}
\usepackage{graphicx}
\usepackage{amsmath}
\usepackage{soul}
\usepackage{float}
\setcitestyle{numbers, sort&compress}

\begin{document}
\title{Simultaneous dynamic characterization of charge and structural motion during ferroelectric switching}
\author{C. Kwamen}
\affiliation{Helmholtz Zentrum Berlin, Albert-Einstein-Str. 15, 12489
  Berlin, Germany}
\author{M. R\"ossle}
\affiliation{Institut f\"ur Physik \& Astronomie,
  Universit\"at Potsdam, Karl-Liebknecht-Str. 24-25, 14476 Potsdam,
  Germany}
\author{M. Reinhardt}
\affiliation{Helmholtz Zentrum Berlin, Albert-Einstein-Str. 15, 12489
  Berlin, Germany}  
  
\author{W. Leitenberger}\author{F. Zamponi}
\affiliation{Institut f\"ur Physik \& Astronomie,
  Universit\"at Potsdam, Karl-Liebknecht-Str. 24-25, 14476 Potsdam,
  Germany}
\author{M. Alexe}
\affiliation{Department of physics,
  University of Warwick, Coventry CV4 7AL}
\author{M. Bargheer} \email{bargheer@uni-potsdam.de}
\homepage{http://www.udkm.physik.uni-potsdam.de} \affiliation{Institut
  f\"ur Physik \& Astronomie, Universit\"at Potsdam,
  Karl-Liebknecht-Str. 24-25, 14476 Potsdam, Germany}
\affiliation{Helmholtz Zentrum Berlin, Albert-Einstein-Str. 15, 12489
  Berlin, Germany}

  \newcommand{\superscript}[1]{\ensuremath{^{\textrm{#1}}}}
\newcommand{\subscript}[1]{\ensuremath{_{\textrm{#1}}}}

\date{\today}
\begin{abstract}
  Monitoring structural changes in ferroelectric thin films during electric field-induced polarization switching is important for a full microscopic understanding of the coupled motion of charges, atoms and domain walls. We combine standard ferroelectric test-cycles with time-resolved x-ray diffraction to investigate the response of a nanoscale ferroelectric oxide capacitor upon charging, discharging and switching. Piezoelectric strain develops during the electronic RC time constant and additionally during structural domain-wall creep. The complex atomic motion during ferroelectric polarization reversal starts with a negative piezoelectric response to the charge flow triggered by voltage pulses. Incomplete screening limits the compressive strain.  The piezoelectric modulation of the unit cell tweaks the energy barrier between the two polarization states. Domain wall motion is evidenced by a broadening of the in-plane components of Bragg reflections. Such simultaneous measurements on a working device elucidate and visualize the complex interplay of charge flow and structural motion and challenges theoretical modelling.
\end{abstract}

\maketitle
\maketitle
The applications of ferroelectric (FE) materials in technology have many facets based on the fundamental aspects of charge transport, structural changes of the crystal lattice and the motion of domain walls. Sensors and actuators mostly rely on the large piezoelectric coefficient $d_{33}$, which can be particularly high e.g. in relaxor ferroelectrics \cite{li2016} or at morphotropic phase boundaries \cite{liu2009}. In electronic circuits the large dielectric constant $\epsilon$ yields a high capacitance $C=\epsilon \epsilon_0 A/d$ in capacitors with area $A$ and thickness $d$. Recently, the concept of negative capacitance has re-attracted attention\cite{khan2015,sala2008}. Considerable  research efforts were devoted to finding the minimum thickness for sustaining ferroelectricity \cite{spal2004}, not only because $d$ is in the denominator of the capacitance, but mainly because of the quest for FE-RAMs with high storage density and low power consumption \cite{junq2003,sett2006,fuji2013}. The switchable FE polarization $P_S$ allows storing an information bit as the sign of $P_S$ of only a few unit cells \cite{junq2003}. In these elements, domain-wall motion is unwanted as it slows down the response due to creep motion \cite{tybe2002}, which is limited by defects and inherently bearing the problem of fatigue. In most materials domains nucleate at defect sites. First, needle-like domains start to grow parallel to the external field and spread out perpendicular to it \cite{scot2010}. Strain tuning of ferroelectric thin films holds a great promise for using various materials in heterostructures with new functionalities \cite{schl2007,shuk2015}. In many more advanced electronic devices, e.g. ferroelectric tunnel junctions, the complex interplay between charge and structural motion becomes important \cite{gaje2007,wen2013,tsym2013,pant2012}.
Many researchers employ piezo-force microscopy to locally switch ferroelectric materials and to detect the strain response to applied fields,
down to the 100 ns time scale \cite{gruv2008}.  Timescales down to 100 fs, resolving the atomic motion of the ferroelectric soft mode, can be accessed by time-resolved x-ray diffraction (tr-XRD) \cite{korf2007b,schi2013a}. The piezoelectric (PE) response\cite{grig2006} and the domain-wall dynamics \cite{grig2006b} were observed on a few nanosecond timescale using microdiffraction. 
Tr-XRD measurements on polycrystalline ferroelastic domain switching on the timescale given by a 1 kHz ferroelectric test sequence showed that resonant x-rays enhance the difference in the structure factor between unit cells with up and down polarization \cite{gorf2016}. In a bulk single crystal, similar experiments revealed that the high piezoelectric response originates from regions where the nucleation of the ferroelectric domains takes place \cite{gorf2015}. For polycrystalline samples the 90$^{\circ}$ ferroelastic switching is observed by two nearby peaks of similar magnitude corresponding to the substantial tetragonality ratio $c/a=1.06$ of Pb(Zr,Ti)O$_3$ (PZT) \cite{dani2014}. Such tr-XRD experiments are important in various fields of materials science, ranging from ceramics \cite{pram2009} to the modeling of devices \cite{mill1990}. The role of structural disorder in relaxor ferroelectrics is accessible with x-rays, too \cite{xu2006}.
For thin films, piezo-force experiments combined with molecular dynamics simulations revealed that in PZT thin films, the prevalent switching mechanism depends on the crystalline orientation \cite{xu2015}. Ultrafast x-ray diffraction experiments following optical excitation revealed the importance of depolarization field screening in the first few ps of the dynamics \cite{dara2012}. Conducting such ultrafast time-resolved experiments for potential laser-assisted ferroelectric switching crucially requires full electronic control for resetting the initial monodomain state in pump-probe experiments, which is possible by the electrical switching scheme presented in the following.

In this study, we use time-resolved reciprocal space mapping (tr-RSM) to measure the structural dynamics in a Pb(Zr$_{0.2}$Ti$_{0.8}$)O$_3$ (PZT) thin-film capacitor while it undergoes a standard ferroelectric test cycle. We selected a large-area capacitor with a rather slow RC time constant of $\tau_{RC}=1$~$\mu$s for charging and discharging.
\begin{figure}[t]
  \centering
  \includegraphics[width=8.6cm]{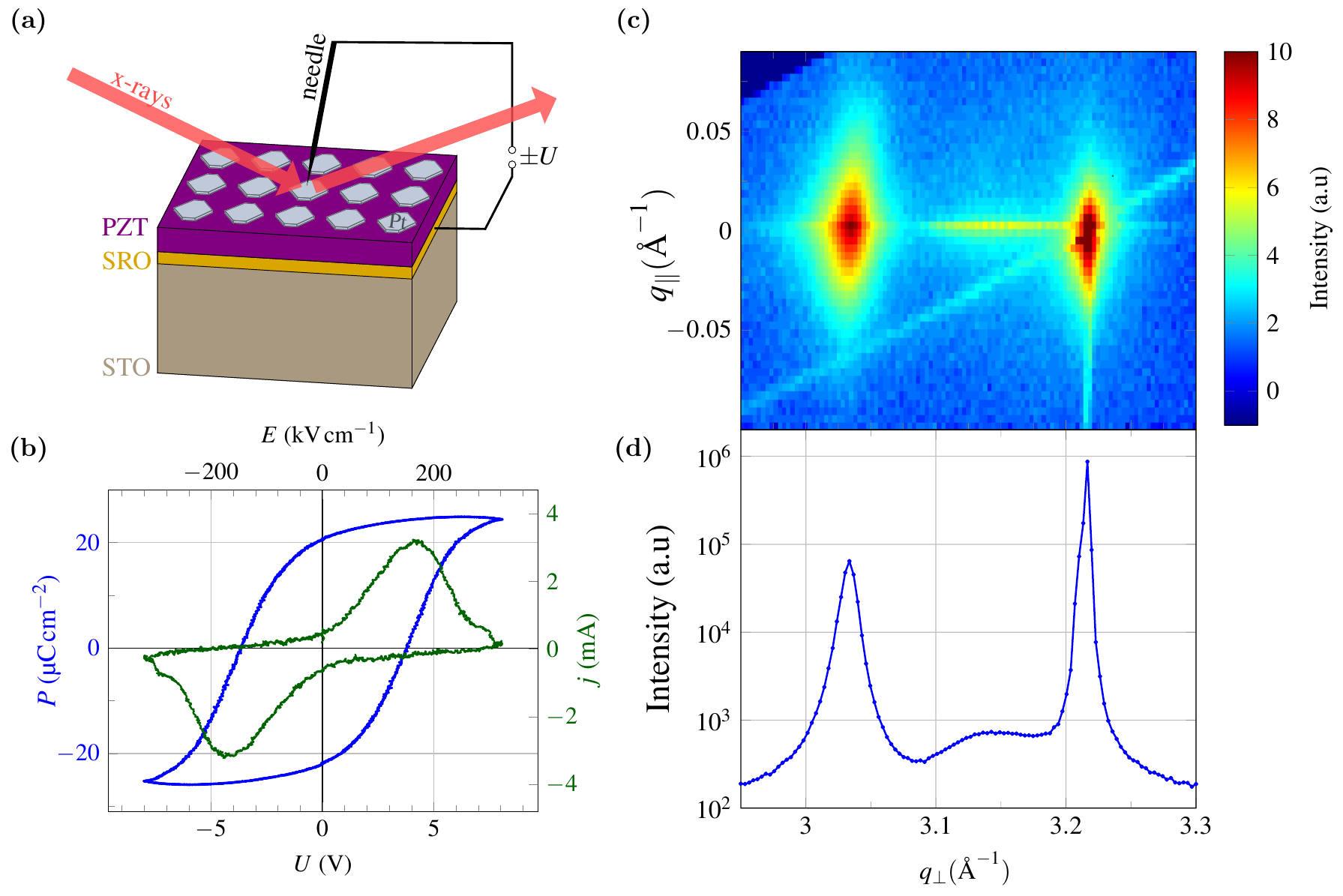}
  \caption{\textbf{Device Characterization} a) Sample structure showing the stacking sequence
    of the different layers , connections to external bias, and direction
    of x-ray beam. b) Polarization-electric field hysteresis loop
    and switching current recorded at 4 kHz. c) typical RSM image around the 002 Bragg reflections of
    PZT, SRO and STO of the as-grown film and d) line profile obtained from panel c) at $q_{||}=0$.
    }
\label{fig:Sample Geometry}

\end{figure}
This allows us to measure the sequence of electrical and structural events taking place during the polarization reversal. Immediately after a current starts charging the capacitor with the opposite polarity, a pronounced negative piezoelectric strain develops within 1~$\mu$s. The ferroelectric switching current connected to the polarization reversal sets in more slowly and reaches its maximum after 4~$\mu$s. We conclude that the screening of the switched ferroelectric charges is incomplete, as the charge flow through the circuit from cathode to anode lags behind by $\tau_{RC}$. This reduces the macroscopic field inside the sample, which is responsible for the negative piezoelectric response, explaining why the positive piezoelectric response is about five times larger.
The tr-RSM additionally reports on the positions of positive and negative ions during the switching, including a characteristic structural disorder within the unit cells and a domain formation and growth.
Our measurements benchmark the x-ray based reciprocal-space analysis of a working FE switch that operates millions of times at several kHz repetition rate. In this paper, we mainly discuss the observation of the functioning switch observed in many devices, whereas the fatigue processes taking typically $10^8$ cycles strongly depend on each electrode and shall be discussed elsewhere.

We report data on a 250 nm lead zirconium titanate (PZT) epitaxially grown on a strontium ruthenate (SRO) electrode with (001) orientation on strontium titanate (STO), i.e. with the elongated $c$-axis of PZT perpendicular to the film. (For details see Methods). Fig.~\ref{fig:Sample Geometry}a) shows the device structure with large Pt top electrodes with 0.5 mm diameter, which permits electrically contacting the film with a needle without shadowing the hard x-ray probe pulse. We limit the applied external fields to $E=8$~V~$/250$~nm~$\approx 320$ kV/cm to avoid the rapid fatigue observed for a bias of about 10 V. Fig.~\ref{fig:Sample Geometry}b) shows the hysteresis loop of the device derived from a triangular bipolar driving voltage at a repetition rate of 4 kHz. The slope of the curve near the coercive field of about 150 kV/cm hints at a spatial variation of the switching response resulting from multiple domains nucleating in the thin epitaxial film.
The reciprocal space map \cite{schi2013a,schi2013d,rein2016} in Fig.~\ref{fig:Sample Geometry}c) shows the scattered x-ray intensities as a function of the wavevector transfer parallel $q_{\parallel}$ and perpendicular $q_{\perp}$ to the film, revealing the 002 reflections of PZT, SRO and STO. The logarithmic intensity of the projection onto $q_{\perp}$ in Fig.~\ref{fig:Sample Geometry}d) shows the sharp substrate peak and the broad crystal truncation rod of SRO next to the PZT peak, which has a similar broadening along $q_{\parallel}$ and $q_{\perp}$. We extract a lattice constant of 0.414 nm. $90^{\circ}$ $a$-domains of PZT would be observed between the SRO and $c$-domain PZT peaks and we estimate their fraction to be below 5$\%$.
We now discuss the structural information deduced from time-resolved reciprocal space maps simultaneously recorded with the electrical characterization via a PUND sequence. The time resolution of the x-ray diffraction experiment was better than 100 ns and can in principle be reduced to 100 ps (see Methods).

\begin{figure}[ht]
  \centering
  \includegraphics[width=7.5cm]{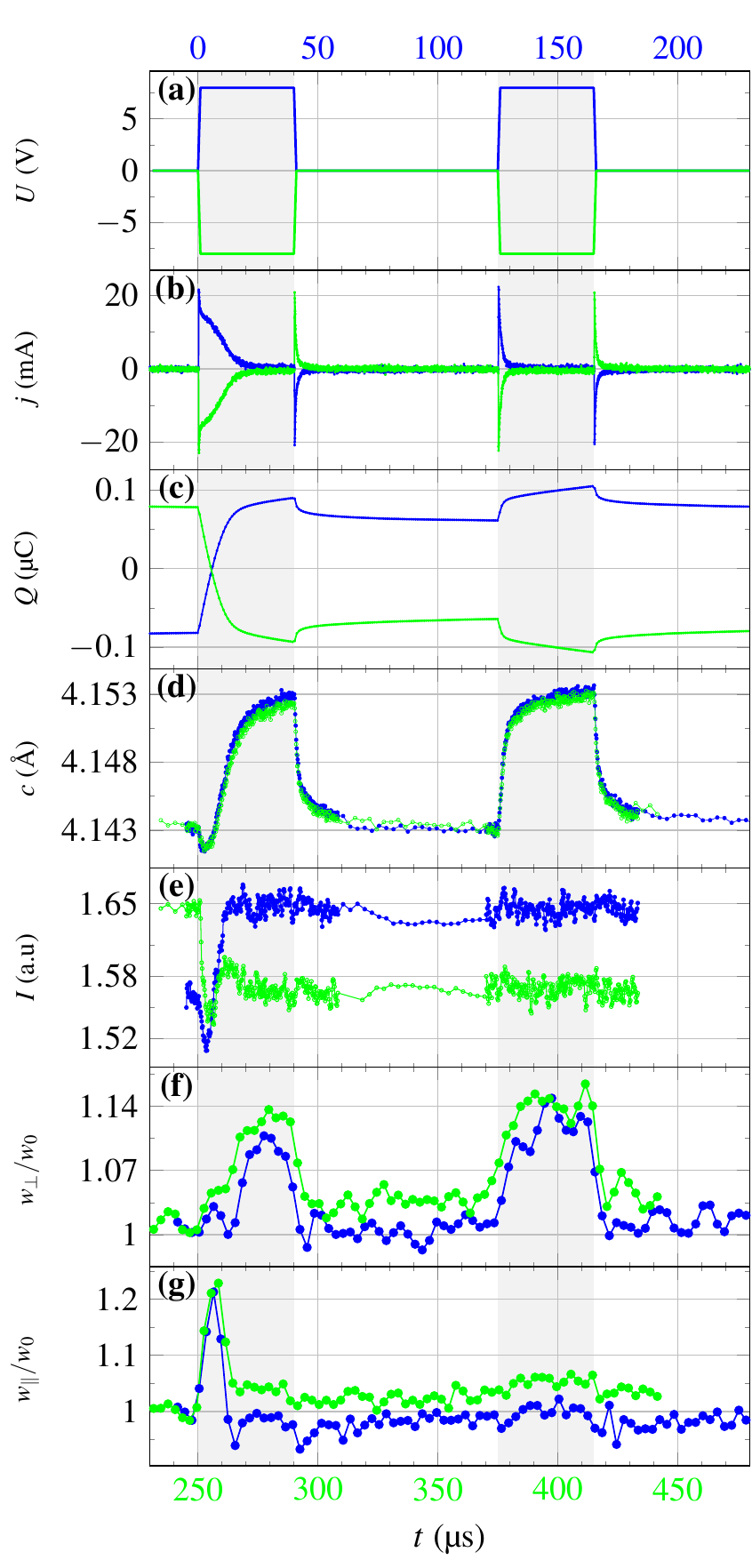}
  \caption{\textbf{Device responses to the PUND sequence.} The blue color represents
    the positive pulses P$_{FE1}$ and P$_{PE1}$ and the green color encodes negative pulses P$_{FE2}$ and P$_{PE2}$.
    a) Applied bias $U(t)$. b) Current response $j(t)$. c) Corresponding charge $Q(t)$ obtained from
    integrating the current signal in b) over time. d) Out-of-plane lattice constant $c(t)$. e)
     integrated  intensity $I(t)$. f) Normalized FWHM of the out-of-plane Bragg
     peak, $w_{\perp}$.  g) Normalized FWHM of the corresponding scattering vector parallel to film
     surface, $w_{\parallel}$.
     }
\label{fig:pund}

\end{figure}

Fig.~\ref{fig:pund} summarizes all the simultaneously measured electrical and structural responses to the applied PUND voltage pulse sequence as a function of time. During this positive-up-negative-down (PUND) pulse sequence, a first pulse (P$_{FE1}$) switches the capacitor, and a second pulse (P$_{PE1}$) with the same magnitude and polarity measures the piezoelectric and capacitive response of the switched sample. The third (P$_{FE2}$) and fourth (P$_{PE2}$) pulse repeat this analysis with opposite polarity.
Fig.~\ref{fig:pund}a) shows the applied voltage pulses. For compactness of the representation in all panels, blue lines encode the positive voltage pulses corresponding to the upper horizontal time axis and green lines encode the negative voltage pulses corresponding to the lower horizontal time axis.

The blue pulses are labeled P$_{FE1}$ and P$_{PE1}$ and the green pulses are labeled P$_{FE2}$ and P$_{PE2}$. The pulses P$_{FE1,2}$ are applied to the sample after a pulse with opposite polarity had been applied. This leads to a ferroelectric polarization reversal. The pulses P$_{PE1,2}$ have the same polarity as the preceding pulse and lead to a charging of the capacitor without switching but they also trigger a piezoelectric response associated with domain wall motion.

The measured current $j$ in Fig.~\ref{fig:pund}b) rises and decays within $\tau_{RC}= 1 \mu$s as a response to the non-switching pulses P$_{PE1,2}$. The response to P$_{FE1,2}$ exhibits an additional current due to FE switching that decays on a time-scale of 10 $\mu$s. We chose the pulse length of 40 $\mu$s and the repetition rate according to the duty cycle of the device. Fig.~\ref{fig:pund}c) shows the charge $Q$ of the capacitor as derived from integration of Fig.~\ref{fig:pund}b). Voltage, current and charge (panels a,b,c) show mirror images for the positive (blue) and negative (green) bias, i.e. the device is essentially symmetric despite the different top- and bottom electrode materials. The time-resolved Bragg-peak shift measured simultaneously, reveals the changes of the out-of-plane lattice constant $c=2\pi / q^{(002)}_{\perp}$ in Fig.~\ref{fig:pund}d), which is essentially the same for the positive and negative pulses. The non-switching pulses P$_{PE1,2}$ yield an expansion of the lattice, whereas just after application of the switching pulses P$_{FE1,2}$, a small compression of the lattice is observed. The integrated peak intensity $I$ in Fig.~\ref{fig:pund}e) shows a clear change of about 5$\%$ upon switching of the device. The peak intensity is larger after switching the device with a positively charged Pt electrode. During the short time, where a lattice compression is observed in Fig.~\ref{fig:pund}d), the peak intensity is transiently suppressed by additional 3$\%$. The peak widths $w_{\perp}$ along $q_{\perp}$, perpendicular to the film (Fig.~\ref{fig:pund}f) and $w_{\parallel}$ along $q_{\parallel}$, parallel to the film plane (Fig.~\ref{fig:pund}g), both increase during the switching pulses P$_{FE1,2}$. However, only $w_{\perp}$ increases during non-switching pulses. $w_{\parallel}$ is only increased in the first 10 $\mu$s of P$_{FE1,2}$, whereas $w_{\perp}$ responds after $w_{\parallel}$ has returned to zero.

We first discuss the capacitive and piezoelectric response of the film upon a non-switching pulse P$_{PE1,2}$. Looking at the charging and discharging current spikes in Fig.~\ref{fig:pund}b), we can fit an exponential decay with the time constant $\tau_{RC}=1\mu$s to the signal. In Fig.~\ref{fig:pund}c) the continuous rise of the charge might at first be attributed to a leakage current, however, Fig.~\ref{fig:pund}d) clearly shows that during this 40~$\mu$s timescale the piezoelectric strain keeps rising. The constant voltage (Fig.~\ref{fig:pund}a) sets the external electric field in the ferroelectric. Therefore the slow additional piezoelectric response which is observed when the external voltage is constant, especially also while the external voltage is zero, must be ascribed to domain wall creep-motion, which accounts for the slow increase and decrease of the charge $Q$ in Fig. 2c. We note that during this process the in-plane peak width Fig.~\ref{fig:pund}f) is constant as this process only switches about 10$\%$ of the polarization, which accounts for the difference in the saturation polarization $P_S$ of the switchable charges and the remanent polarization $P_R$. The increasing out-of plane peak width illustrates inhomogeneities in the piezoelectric response across the large electrode area \cite{grig2006}.
 \begin{figure}[t]
  \centering
  \includegraphics[width=8.5cm]{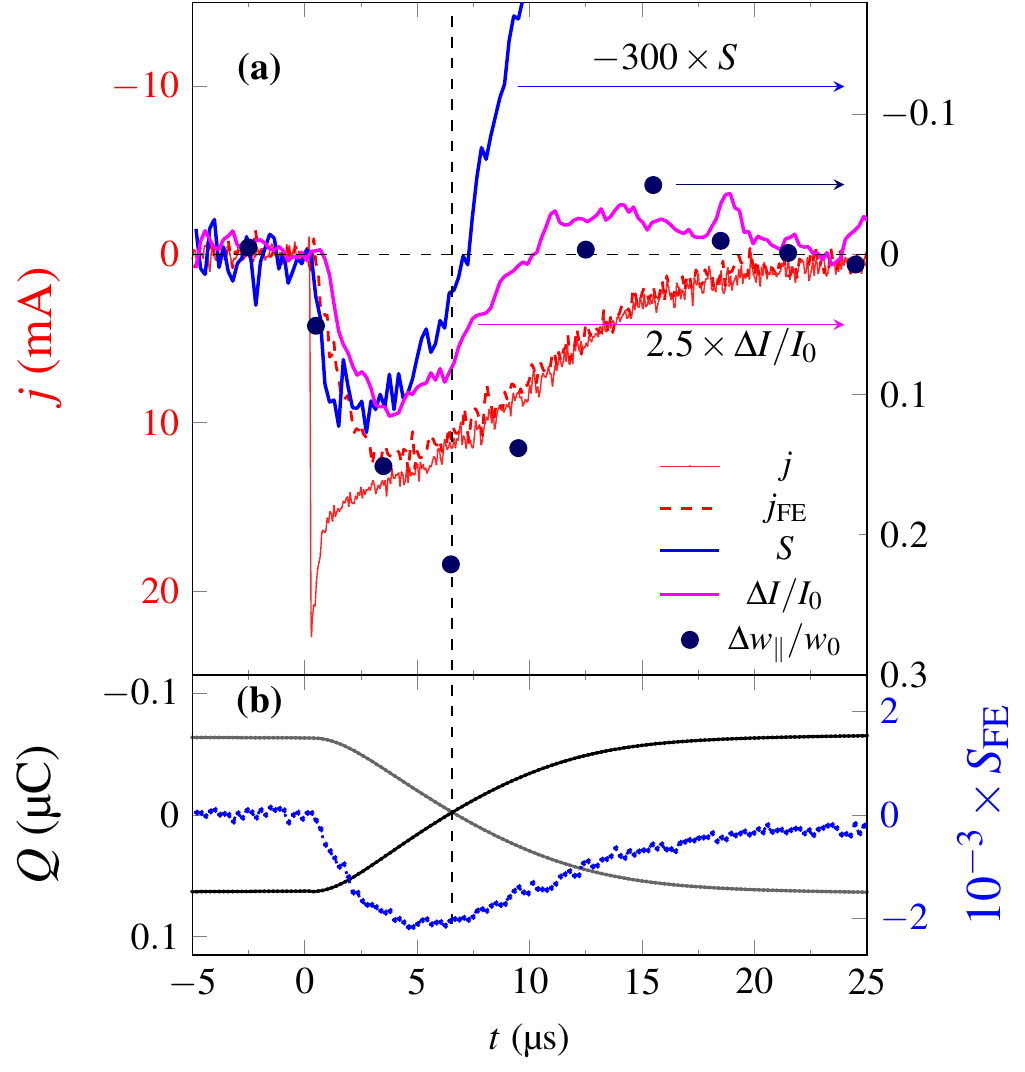}
  \caption{\textbf{Sequence of events during polarization reversal} a) The current (solid red line) starts immediately after the rising edge of the voltage pulse P$_{FE1,2}$. The strain (black line) responds within 1 $\mu$s, whereas the FE part of the current (red dashed) rises later. The violet line shows the drop of the Bragg peak intensity (sum of the responses to P$_{FE1,2}$), which is a measure of the structural disorder. The black dots represent measurements of the in-plane peak width, which is maximum, when half of the thin film is switched (high in-plane domain density). b) Time-evolution of the switchable charges during P$_{FE1,2}$ obtained by subtracting the PE response at P$_{PE1,2}$ from the total charges (Fig.~\ref{fig:pund}c) during the switching. Similar for the FE component of the strain (dashed blue).}
\label{fig:D}

\end{figure}

The structural response during the polarization reversal (P$_{FE1,2}$) has microscopic and mesoscopic components. The peak intensity (Fig.~\ref{fig:pund}e) is higher when the positive bias is applied to the Pt top electrode. According to the commonly accepted structural model, the Ti atom is guided by the O octahedra in the same direction, i.e. the positively charged Ti also moves towards the positive electrode, albeit less than the O. We expect about 7$\%$ structure factor change for the switching from 002 to 00$\bar 2$ due to the violation of Friedel's law \cite{als2011,gorf2016}. The higher intensity is calculated, when the Oxygen octahedra move towards the positively charged Pt electrode through which the x-rays are incident. This static structural model is confirmed by our data for times $t>10$~$\mu$s, when the switching is essentially completed. For $t<10$~$\mu$s both the green and the blue curves in Fig.~\ref{fig:pund}e) show an additional decrease in intensity by 3$\%$. This may be ascribed to a transient structural disorder in the position of the Ti atom and the O octahedra analogous to the thermal disorder described by the Debye-Waller effect.
During the same time, the in-plane peak width $w_{\parallel}$ transiently increases. This must be attributed to a loss of in-plane coherence of the lattice, i.e. domain formation. When $w_{\parallel}$ relaxes around $t=10$~$\mu$s, the out-of-plane peak width
$w_{\perp}$ rises to the same value as during the non-switching pulse. Therefore, the changes $\Delta w_{\perp}$ in the out-of-plane peak width are attributed to different peak strain values attained across an inhomogeneous large electrode. Below $t<10$~$\mu$s, the peak width increase $\Delta w_{\perp}$ is not yet fully developed. At first one would expect a maximum difference in the expansive and contractive strain for the two domain orientations which exist during the switching, because they have opposite piezoelectric effects. We see two reasons for our unexpected observation of an initially reduced $\Delta w_{\perp}$: structural clamping of domains and insufficient screening. As long as the needle-like domains emerging from nucleation sites are laterally  small, the lattice constant is clamped to the adjacent domain that experiences a strain with opposite sign. For small domain size the oppositely poled domains hinder each other's expansion. Secondly, during the ferroelectric switching, the screening of the switched ferroelectric dipoles by charges on the electrode lags behind by $\tau_{RC}$, effectively reducing the field inside the capacitor, as we will see discuss below.

In order to confirm the incomplete screening of the ferroelectric charges during the switching, Fig. \ref{fig:D} displays the simultaneously collected information during the polarization reversal. The current (solid red line in Fig. \ref{fig:D}a) rises immediately after the application of the external voltage. As soon as the charges appear on the electrodes, the electric field in the capacitor equals $E = U/d$. Since in the beginning of the switching pulse P$_{FE1,2}$ the elementary dipoles of the ferroelectric are pointing in the opposite direction, a negative strain (black line) develops at the same rate at which the positive piezoelectric effect rises during P$_{PE1,2}$. However, this compression already stops after $1\mu$s, although only a very small fraction of less than 5$\%$ of the crystal has been switched, according to the measured charge (\ref{fig:D}b). The screening of these switched ferroelectric charges lags behind. This partially cancels the applied electric field, explaining why the amplitude of the compression is 5 times smaller than the following expansion. This is consistent with the reduction of the Bragg peak intensity, as the weaker field imposes a less strict ordering of the atomic positions. The green line indicating the reduction of the peak intensity $\Delta I/I_0 = I_{P_{FE1}}+I_{P_{FE2}}$ due to disorder is obtained by adding the intensities measured during P$_{FE1}$ and P$_{FE2}$. The switching current $j_{FE}=j_{tot}-j_{PE}$ induced by the polarization reversal (dashed red) in the ferroelectric is obtained by subtracting the piezoelectric response during P$_{PE1}$ from the ferroelectric response P$_{FE1}$. Note that the onset of the disorder and the onset of the ferroelectric switching current lag behind the piezoelectric contraction. In other words, the disorder and the ferroelectric switching current proceed, while the potential barrier between both polarization states is suppressed by the compressive strain in the domains that are not yet switched \cite{cohe1992}. For the observed compressive strain of $\Delta c_{min}/c_0=4\cdot10^{-4}$ and expansive strain $\Delta c_{max}/c_0=2\cdot10^{-3}$ we estimate the modulation of the potential barrier to be 1 to 5 meV, based on ab-initio calculations, which predict a barrier suppression from 0.23 eV for $c/a=1.06$ to 0.1 eV for $c/a=1$ \cite{cohe1992}.
\begin{figure}[t]
  \centering
  \includegraphics[width=8.5cm]{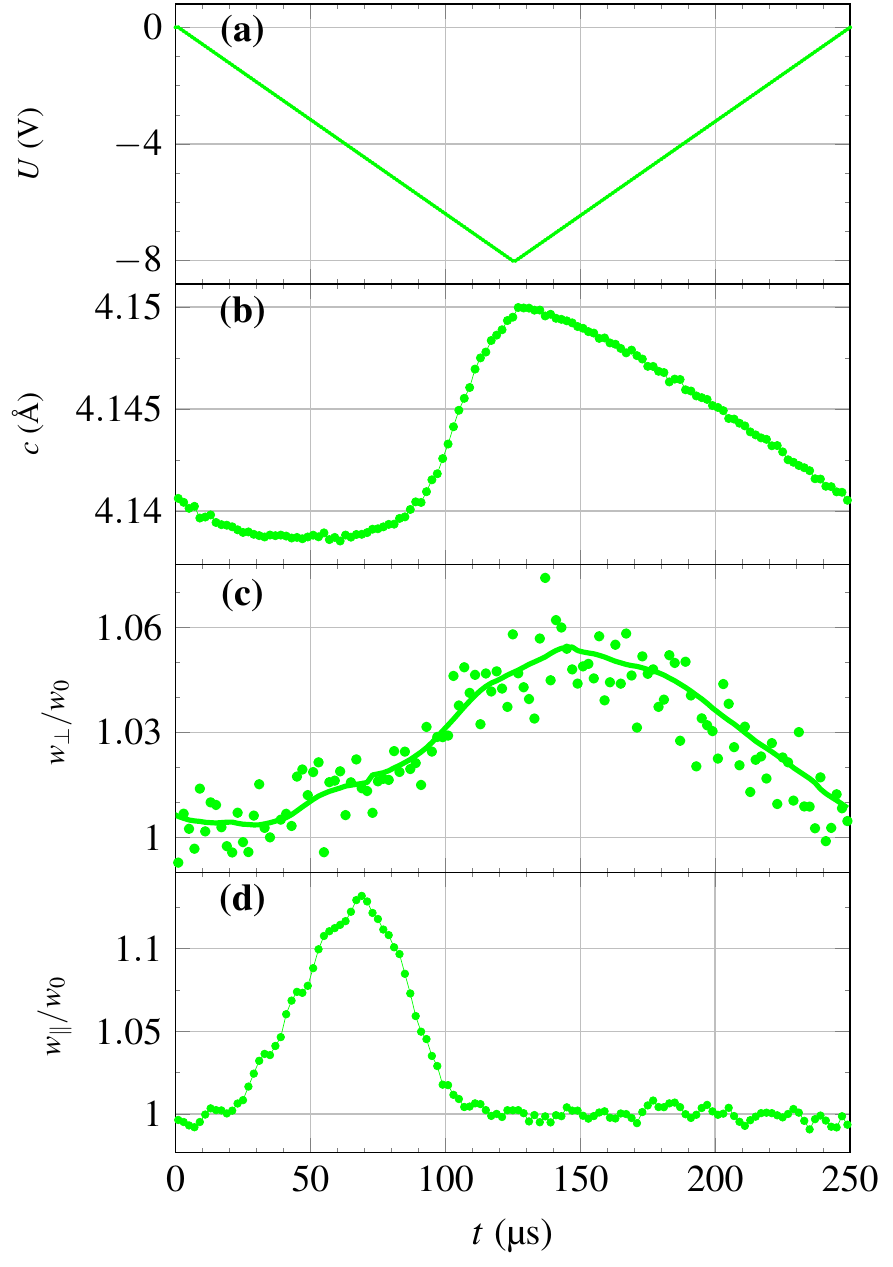}
  \caption{\textbf{Switching with triangular pulse.} a) Applied voltage pulse pattern $U(t)$.
    b) Out-of-plane lattice constant $c(t)$. c) Normalized FWHM of the out-of-plane Bragg peak
    and d) Normalized FWHM of the corresponding lateral peak.
    }
\label{fig:tri}

\end{figure}

Fig. \ref{fig:D}b) shows the charge redistribution for the positive and negative ferroelectric parts of the  test cycle from an integration of $j_{FE}$. The two lines cross at the time $\tau_{1/2}$, when half of the sample has been switched, and the total strain (black line in \ref{fig:D}a) crosses through zero strain.

In order to confirm our conclusions and to highlight the timing of the structural response during the switching, we carefully investigated the effect of a negative triangular voltage ramp, which is half of the cycle used to measure the hysteresis loop in Fig.~\ref{fig:Sample Geometry}b). Starting at the pulse onset $t=0$, the lattice constant shows a small gradual contraction confirming the negative PE response in the initial stage of the switching. When the bias in the film reaches $U=3.7$~V, i.e. the coercive field of $E\approx 150$ kV/cm, the lattice constant reaches its minimum and subsequently shows a rapid and strong expansion due to the positive PE effect. The maximum negative piezoelectric response $\Delta c_{min}$ is larger in this triangular pulse sequence compared to the square-voltage pulses. At first, one would expect it to be smaller, because the applied voltage is smaller ($U=3.7$V instead of 8 V) at the time $t_{min}$, where the minimum occurs. In addition, the slowly rising field should give the film more time to reverse the polarization. We conclude that the fast application of the external field leads to a depolarization field associated with the switched domains, which is not yet fully screened because the charges on the electrode lag behind by $\tau_{RC}$, explaining the rather small negative piezoelectric response.
The peak width increase along $q_{\perp}$ mimics the response of the PE expansion after  the coercive field is surpassed. In contrast, the peak width along $q_{\parallel}$ increases already at very small fields, reaches a maximum near the coercive field and then rapidly decreases as the voltage is increased.

In conclusion, we have simultaneously observed the charge flow and the structural motion during FE polarization reversal. We find piezoelectric charge flow connected to domain wall creep, and field induced switching that starts with a negative piezoelectric response. It is about five times smaller than the positive piezoelectric strain observed after the switching, because during the switching, the depolarization field of the switched dipoles is not fully screened by external charges on the electrodes. The piezo-response modulates the energy barrier between up and down polarization states. We believe that the experimental results found by simultaneous assessment of the structural and electronic response of this prototype ferroelectric capacitor will be useful in the devising new concepts for ferroelectric polarization reversal.

\textbf{Methods}
The investigated device  consists of metal-ferroelectric-metal capacitors based on an epitaxially grown 250 nm Pb(Zr$_{0.2}$Ti$_{0.8}$O$_3$ (PZT) film  on a 50 nm thick SrRuO$_3$ (SRO) bottom electrode deposited by pulsed laser deposition on a (001) oriented SrTiO$_3$ (STO) substrate \cite{vrej2006}. Fig.~\ref{fig:Sample Geometry}a) schematically shows this structure with 20 nm thick hexagonal Pt top electrodes having 0.3 mm edge length and 0.5 mm diameter.

An external electric field was applied to individual Pt top electrodes of the device via a biased needle. The function generator model HP33120A applied voltages up to $U =\pm 8$ V, well below the damage threshold of about 10 V and the resulting current signal was measured indirectly via the $50\,\Omega$ resistance of an oscilloscope (Agilent).

The X-ray experiment is performed at the XPP-KMC3 beamline at the synchrotron-radiation facility BESSY II of the Helmholtz-Zentrum Berlin.  A 200 $\mu$m wide x-ray beam at 9000 eV photon energy was aligned to a single Pt electrode in Bragg-geometry via a 4 circle goniometer. The reciprocal space map \cite{schi2013a,schi2013d} in Fig.~\ref{fig:Sample Geometry}c) is recorded by collecting the symmetrically and asymmetrically scattered x-rays by a two-dimensional pixel detector (Pilatus 100k, Dectrics) for various angles $\omega$ between the incoming x-rays and the sample surface. A unitary transformation maps this signal onto the scattering intensities \cite{schi2013a,schi2013d} shown as a function of $q_{\parallel}$ and $q_{\perp}$ in Fig.~\ref{fig:Sample Geometry}c).

In order to combine the electrical PUND-analysis with time resolved x-ray measurements 
we synchronized the gate of the Pilatus detector to the rising edge of the electric pulse. An electronic timing allows variation of the delay between the detector gate and the onset of the pulse-sequence. For a fully time-resolved reciprocal space mapping, the time resolution is limited by the smallest possible window of the detector which is about 100 ns. Since this method only uses a twentieth of the full x-ray flux, some measurements are carried out with a point-detector realized by combining a fast x-ray scintillator with a photomultiplier tube read out in single-photon counting mode \cite{navi2012a}. The time-resolution is then given by the 2 ns response time of the scintillator and we can now continuously measure the full pulse sequence with all x-rays emitted at the beamline, since each photon is time-correlated with the applied electrical pulse sequence using a current amplifier (FEMTO) and the  time-correlated single-photon counting  module PicoHarp 300 (PicoQuant). In any case, the time resolution is much faster than the response time of the FE device.

\textbf{Acknowledgement:}
MA acknowledges the financial support of Royal Society through Wolfson Research Merit and Theo Murphy Blue Sky Awards. CK would like to thank Q. Cui and W. Wirges for help with the ferroelectric test cycles.
\bibstyle{unsrt}

\end{document}